\documentclass[12pt,preprint]{aastex}

\newcommand{\Msun}{\mbox{$\cal M_{\odot}$}}

\newcommand{\fuvmag}{\mbox{\it FUV~}}
\newcommand{\nuvmag}{\mbox{\it NUV~}}
\shortauthors{Bianchi et al.}
\shorttitle{Star formation in M101 and M51 from GALEX Imaging}
\begin{document}
\title{Recent star formation in nearby galaxies from GALEX imaging:
   M101 and M51}
\author{
Luciana Bianchi\altaffilmark{1}, David A. Thilker\altaffilmark{1},
Denis Burgarella\altaffilmark{4}, 
Peter G. Friedman\altaffilmark{2},
Charles G. Hoopes\altaffilmark{1},
Samuel Boissier\altaffilmark{7}, Armando Gil de Paz\altaffilmark{7},
Tom A. Barlow\altaffilmark{2}, 
Yong-Ik Byun\altaffilmark{3}, Jose Donas\altaffilmark{4},
Karl Forster\altaffilmark{2},
Timothy M. Heckman\altaffilmark{1}, Patrick N. Jelinsky\altaffilmark{6},
Young-Wook  Lee\altaffilmark{3}, Barry F. Madore\altaffilmark{7},
Roger F.  Malina\altaffilmark{4},
D. Christopher Martin\altaffilmark{2}, Bruno Milliard\altaffilmark{4},
Patrick  Morrissey\altaffilmark{2}, Susan G. Neff\altaffilmark{9},
R. Michael Rich\altaffilmark{10}, David Schiminovich\altaffilmark{2},
Oswald H. W.  Siegmund\altaffilmark{6}, Todd Small\altaffilmark{2},
Alex S. Szalay\altaffilmark{1}, Barry Y. Welsh\altaffilmark{6},
Ted K. Wyder\altaffilmark{2}
}
\altaffiltext{1}{
Deptartment of Phys.\& Astron., Johns Hopkins
University,3400 N.Charles St., Baltimore, MD21218
(bianchi,dthilker,hoopes,heckman@pha.jhu.edu)}
\altaffiltext{2}{California~Inst. of Technology,MC405-47,1200 E.California Blvd, Pasadena, CA91125 
(friedman,tab,krl,cmartin,patrick,ds,tas,wyder@srl.caltech.edu)}
\altaffiltext{3}{Center for Space Astrophysics, Yonsei University, Seoul
120-749, Korea
(byun,ywlee@obs.yonsei.ac.kr)}
\altaffiltext{4}{Laboratoire d'Astrophysique de
Marseille, BP8, Traverse du Siphon, 13376 Marseille Cedex 12,FR
(denis.burgarella,roger.malina,bruno.milliard@oamp.fr)}
\altaffiltext{6}{Space Sciences Laboratory, University of California at
Berkeley, 601 Campbell Hall, Berkeley, CA 94720
(patj,ossy,bwelsh@ssl.berkeley.edu)}
\altaffiltext{7}{Observatories of the Carnegie Institution of Washington,
813 Santa Barbara St., Pasadena, CA 91101
(agpaz,barry@ipac.caltech.edu)}
\altaffiltext{9}{Laboratory for Astronomy and Solar Physics, NASA Goddard
Space Flight Center, Greenbelt, MD 20771
(neff@stars.gsfc.nasa.gov)}
\altaffiltext{10}{Department of Physics and Astronomy, University of
California, Los Angeles, CA 90095
(rmr@astro.ucla.edu)}

\begin{abstract}

The GALEX (Galaxy Evolution Explorer) Nearby Galaxies Survey  
is providing deep far-UV and near-UV imaging for a representative
sample of galaxies in the local universe.  We present early results
for M51 and M101, from GALEX UV imaging and SDSS optical data in five bands.
The multi-band photometry of compact stellar complexes in  M101
is compared to  population synthesis models, 
to derive ages, reddening, reddening-corrected luminosities
and current/initial  masses.
The GALEX UV photometry  provides a complete census of young compact 
complexes on a $\approx$ 160~pc scale. 
A galactocentric gradient of the far-UV~-~near-UV color indicates younger
stellar populations towards the outer parts of the  galaxy disks, the effect being
more pronounced in M101 than in M51.

\end{abstract}





\section{Introduction}
\label{sintro}

The study of stellar cluster systems and compact associations provides
insight into a galaxy's star-formation history and stellar content. 
Far-UV and near-UV bands are a sensitive probe 
for detection of young stellar clusters, and for measurement of
 their physical parameters and - combined with optical data - 
of their extinction (Bianchi et al. 1999, 
Chandar et al. 1999).

The GALEX Nearby Galaxies Survey (NGS), described by  Bianchi et al. (2004a, b)
is providing deep far-UV (FUV) and near-UV (NUV) imaging for a representative
sample of $\approx$ 200 galaxies in the local universe. 
 We present early GALEX NGS  data of  M101 and M51. We also use
five-band optical imaging  from the Sloan Digital Sky Survey (SDSS). 
We derive ages, reddening, and current/initial  masses for the compact 
UV sources in M101 (Sect. \ref{sfits}).
Radial profiles of UV brightness and color are presented
in Sect. \ref{sprofiles}.  

The two galaxies studied in this paper have been previously imaged in 
a broad far-UV band ($\lambda$$_{eff}$ $\approx$ 1521\AA) 
by the {\it Ultraviolet Imaging Telescope} (UIT). The UIT data were analysed
by Kuchinski et al. (2000), Waller et al. 1997 (M101), Hill et al. 1997 (M51),
Hoopes, Walterbos, \& Bothun (2001).
While the UIT FUV images revealed morphological differences with respect
to the optical bands, and the brightest stellar complexes, the
GALEX data yield significant advances. First, both FUV and NUV bands
are available, providing an essentially reddening-free color 
which gives a direct indication of the age (for compact sources)
independent of the extinction
estimate, while the UV-to-optical wide baseline is very sensitive to both 
extinction and age (Sect. 3). Second, GALEX's detectors afford  greater
sensitivity and linearity than the UIT data (which were recorded on film),
 measuring UV emission out to larger disk radii, and down to
fainter (by about 3 mag) fluxes (Sect. 4).

\section{Observations and Data Analysis}

The GALEX instrument is described by Martin et al. (2004, this issue)
and its on-orbit performance by Morrissey et al. (2004, this issue). 
GALEX FUV (1350--1750~\AA) and NUV (1750--2750~\AA) imaging
 was obtained with 
total exposure times of 1414 and 1041 sec for M51 and M101. 
Assuming distances of 9.6 Mpc,
and 7.4 Mpc (
M51: Sandage \& Tammann 1974, M101: Kelson et al. 1996) the $4.6\arcsec$ ~FWHM GALEX
PSF corresponds to 209, and 157 pc. 
The 1-$\sigma$ (at the PSF scale) NUV(FUV) sensitivity limit 
of the GALEX images is 27.9 (27.8) AB mag  \arcsec$^{-2}$
for M51 and 27.6 (27.5) AB mag \arcsec$^{-2}$ for M101.

\begin{figure}
\plotone{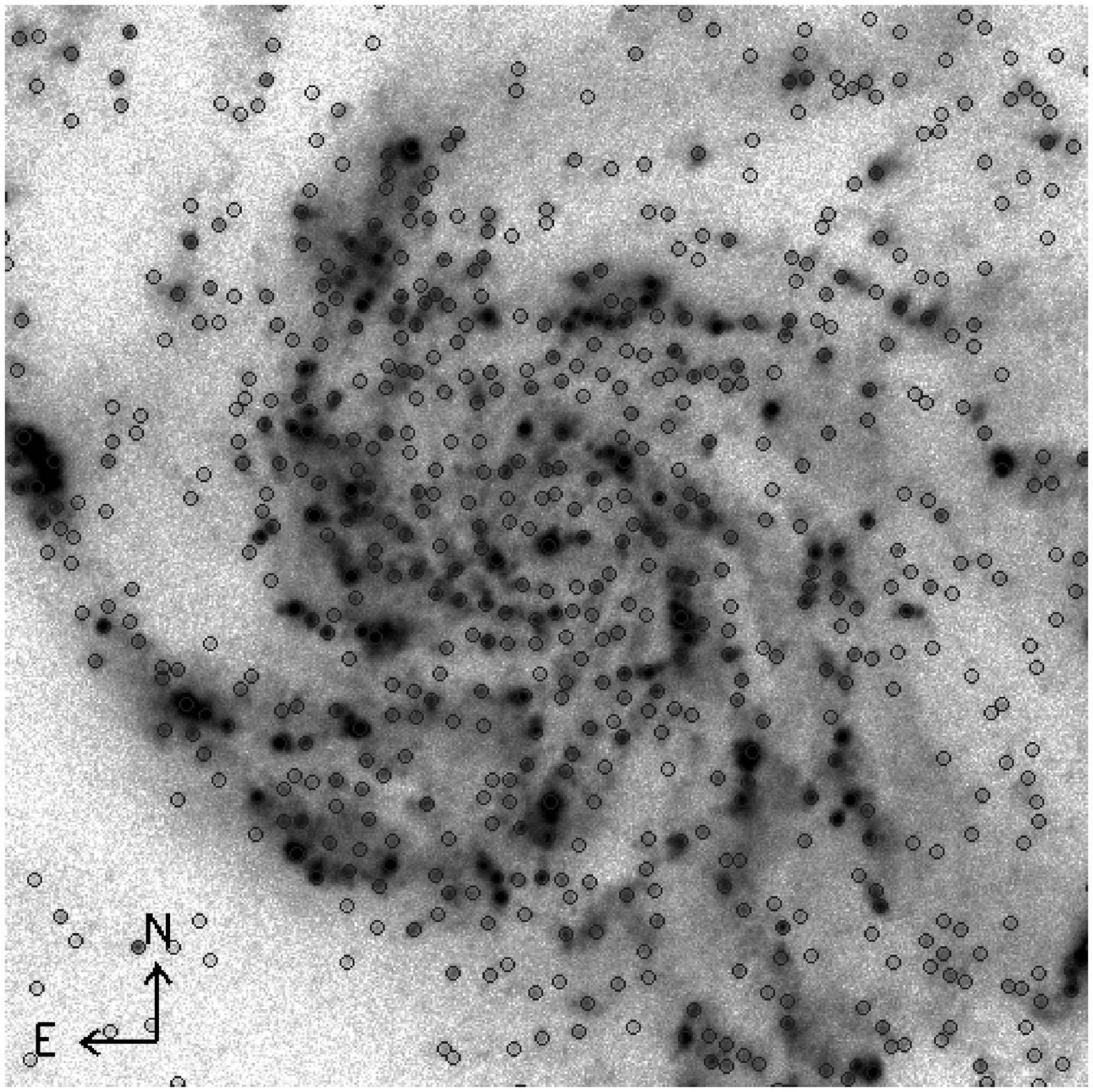} 
\caption{A  portion (12.5$'$ square) of a GALEX-NUV 
image of M101.  The compact UV 
sources with  photometric errors $<$0.2 mag are indicated
with  circles. 
\label{foverlay101} }
\end{figure}

\begin{figure}
\plotone{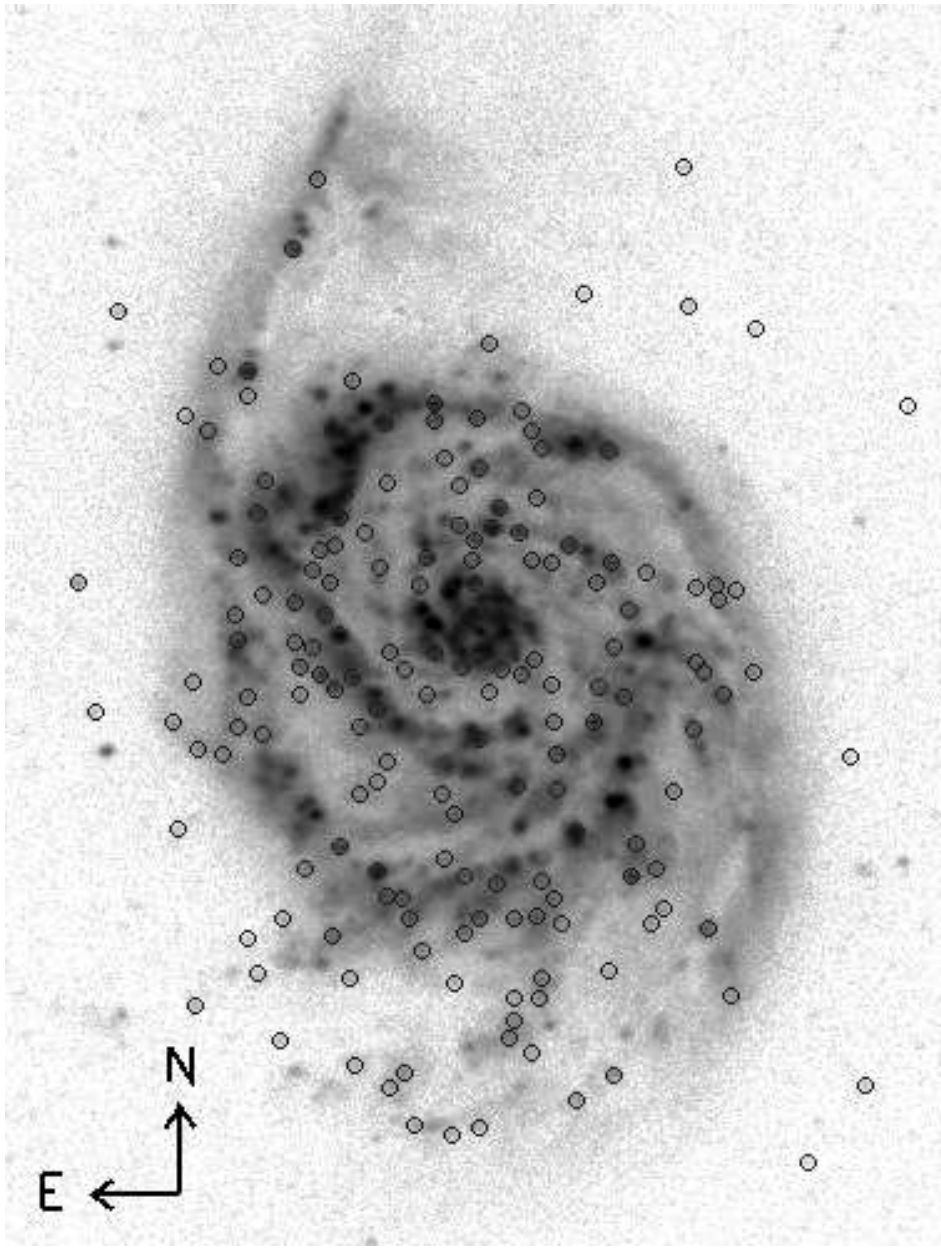}  
\caption{Same as Fig. \ref{foverlay101}, for M51 (9.5x12.5$'$). 
\label{foverlay51} }
\end{figure}

We also  used
SDSS imaging of M51 and M101 in the {\it ugriz} filters. 
The resolution of the optical data (1-2$''$,  $\approx$ 
70 and 50 pc for M51 and M101 respectively) is superior to that of GALEX, 
and in some cases the GALEX UV sources are
resolved into two or three optical sources. 
The different angular resolutions were taken into account
when matching the UV and optical sources. 
 For the analysis in Sect. 3, we focus on the smallest scale detections in each galaxy,
thus sampling the 
 star forming complexes in M101 and M51 on a scale corresponding to the GALEX resolution. 
Post-pipeline processing and photometry on the GALEX and SDSS images
were performed using the SEXTRACTOR (Bertin \& Arnouts 1996) and IRAF APPHOT packages.  To
obtain our master list of ``point-like'' sources we first
ran SEXTRACTOR on the NUV imagery after removal of the diffuse background,
using a detection kernel matched to the GALEX NUV PSF (4.6$''$).  The
background was estimated using a circular median-filter of diameter
10.5$''$ (7 pixels).  Such background removal is
crucial in order to obtain a complete SEXTRACTOR catalog in bright
regions of M51 and M101.  We then passed this list of compact NUV
sources to the APPHOT package, running PHOT  (with recentering enabled)
on background subtracted versions of the GALEX FUV, NUV, and convolved
SDSS images.  Sources which shifted in position by more than 4$''$ in
any band were eliminated.  Finally, PHOT was used to measure fluxes 
for all surviving compact sources with
4$''$ aperture radius, and  aperture corrections based on isolated field
stars were applied. 
 Figures \ref{foverlay101} and \ref{foverlay51} show the GALEX  NUV image 
for the two galaxies.  The matched sources with photometric errors
better than 0.2~mag in the UV bands are shown. 
Table \ref{tlimits} gives the number of detected sources in each band and the statistics of sources
for given error limits.

The final matched catalog was based on the NUV source list 
(the deepest UV band), and  
includes only  NUV sources having a counterpart in 
at least one 
other band.
The photometry procedures were verified by comparing
our results with both GALEX and SDSS 
pipeline photometry on portions of the same images
outside the galaxy body, where the standard pipeline results are reliable.
We found complete consistency within the errors. We applied our
photometry technique in the regions containing the galaxies, where the 
standard GALEX and SDSS pipelines
break down. We use the flux calibration from the GALEX Internal
Release version 0.2. 

\section{The physical parameters of the compact complexes}
\label{sfits}

The measured colors for the compact UV sources in M101  were 
compared to synthetic colors 
computed from the Bruzual and Charlot (2003) models for Integrated
 Populations, for the cases of Single Burst (hereafter SSP)  and of
Continuous Star Formation (hereafter CSP),
as a function of age. As expected (because we are sampling UV-bright compact complexes)
most sources are better represented by a SSP  SED. 
 In this work we assume solar metallicity and
a Chabrier (2003) IMF (mass range 0.1 - 100 \Msun ).  
Because the UV sources are mostly very young complexes, the analysis is
not very sensitive to metallicity.
For instance, the  \fuvmag-\nuvmag  color of a SSP model with solar metallicity
for 1, 10, and 100 Myrs is \fuvmag-\nuvmag = -0.35, -0.06, 0.18 (in the AB mag system).
The  model  colors for the same ages at
z=0.008 differ by  $\le$0.1, which leads to an uncertainty 
in the age by up to  a factor of 3, and typically by $\le$2, depending on the age. 
According to the HII regions  study by Kennicutt, Bresolin, \& Garnett (2003), solar metallicity
is appropriate out to $\approx$6$'$, which includes the majority of our sources.

\begin{figure}
\plotone{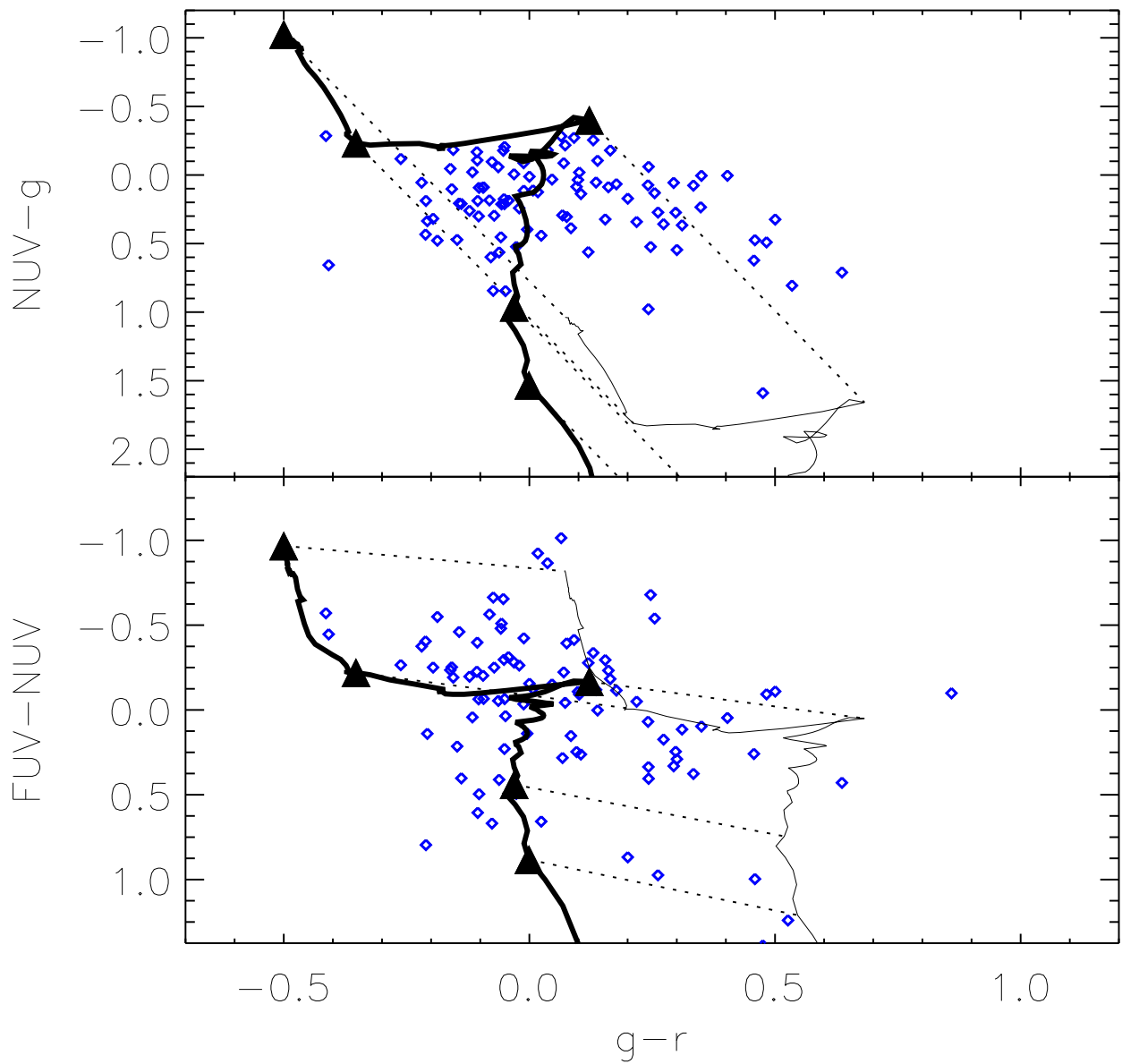} 
\vskip -3.cm
\caption{Color-color diagrams for the compact 
UV sources in M101. Only sources with photometric errors 
$<$0.1 mag (GALEX bands) and $<$0.05 mag (SDSS bands) are plotted. 
The solid 
black curve shows the intrinsic colors for SSP 
models as a function of age. 
Model colors reddened with E(B-V)=0.5
are shown with a thin black line (dotted lines connect the intrinsic
and reddened color for the same model at representative ages). 
Filled black triangles mark ages (in log years) of 6.0, 6.7, 6.9,  8.0,  8.3.}
\label{fcolcol}
\end{figure}

Figure \ref{fcolcol}  shows color-color diagrams of the  sources compared to the models. 
We derived physical parameters by comparing the
measured magnitudes 
to the model colors. Because the
observed SED's depend on the cluster age (intrinsic SED) and 
on the wavelength-dependent interstellar extinction, we reddened the
model colors for different amounts of extinction and  reddening
types. Two  different methods were used, and the results compared.
In the first method, we used the GALEX measurements alone, and estimated
the cluster ages by comparing the \fuvmag-\nuvmag color to the models.
Reddening has a negligible effect on the \fuvmag-\nuvmag color, 
for E(B-V) values  associated with these galaxies.
For instance, model colors for E(B-V)=0.2 (MW-type extinction, R$_V$=3.1) 
 differ from the unreddened colors by $\le$ 0.01, up to 100Myrs. 
In the second method, we  fit all the  measured multi-band magnitudes
 with the synthetic colors to derive age and E(B-V) concurrently,
by $\chi$$^2$ minimization, taking into account the photometric errors
in each band (see Bianchi et al. 2004 in prep., for more details).

\begin{figure}
\epsscale{1.0}
\plotone{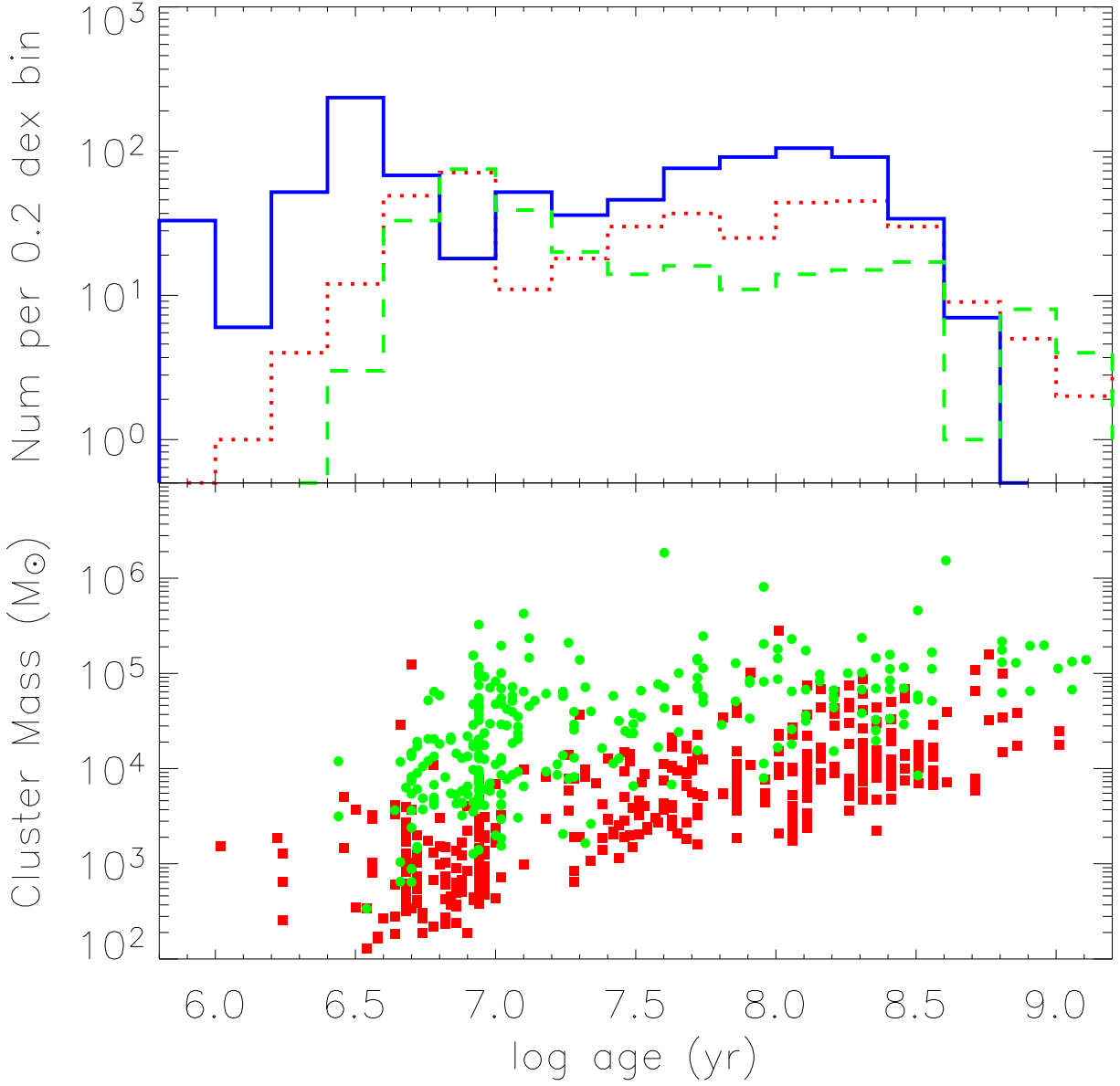}
\vskip -3.cm
\caption{Top: histogram of the derived ages for the
compact stellar complexes in M101 from SED fits (dashed green)
and from \fuvmag-\nuvmag color only (solid blue).
Only sources with small photometric errors {\it} and good SED fits
are included in the green sample, limiting the sample to about 300 sources.
All sources with errors $<$0.2 mag in  \fuvmag and \nuvmag  bands are included
in the blue (solid) histogram (about 1100 sources).
The truncation at ages $<$10$^9$ yrs for the blue histogram (\fuvmag-\nuvmag) is a
selection effect of the sample.
The sharp decline of the green histogram 
at older ages includes selection effects of the sample,
as well as a possible effect of actual disruption.
 The dotted red histogram shows
 the cluster sample from one HST  WFPC2 image
(from Chandar et al. 2004). Bottom: 
masses versus ages derived for our UV sources 
(green dots, parameters from SED fits) and for the cluster
sample of Chandar et al. (2004) (red squares).
\label{fage} }
\end{figure}

Ages determined from the  \fuvmag-\nuvmag color, and by the
SED fit of the UV-to-optical bands respectively, agree for
most of the sources within the errors, but are discrepant for some sources,
where a color gradient may be present. We adopt the age from the SED fits,
which represents the average over the area included in the aperture,
for the sources with  errors $<$0.05-0.10~mag (SDSS and GALEX
bands respectively). These ages are shown in the green-dashed histogram in
Figure \ref{fage}, and
ages derived  by the \fuvmag-\nuvmag  color only (for the  UV sources 
with \fuvmag  and \nuvmag  errors $<$0.2 mag)
are shown by the blue (solid) histogram. This sample (about 1100 sources)
includes a large number of (mostly younger) sources 
eliminated by the error cuts in the SED-fit sample. We note
that individual errors in the ages derived by the \fuvmag-\nuvmag  color in the  range
of log age (in yrs) =6.5-7.5 may be larger than the bin-size, as \fuvmag-\nuvmag
does not vary appreciably in this range.  For this reason, and for the selection 
effects limiting the sources with good SED fits, we do not attempt to
interpret this histogram in terms of star-formation history.
We point out however the great sensitivity of the GALEX UV imaging to 
detect and measure the youngest sources. 
The sharp decline in the number of
sources at older ages is mostly due to our magnitude limits and
color selection effect. However, disruption 
may be  more effective for these large-scale complexes than for bound clusters. 
Once reddening and age are determined for each complex, the 
reddening-corrected luminosity and age yield the current (and thus the
initial)  mass, by comparison with the same models (Figure  \ref{fage}, lower panel).
Figure  \ref{fage} also shows  the cluster
sample detected in one HST WFPC2 field with $U,B,V,I$ measurements
by Chandar et al. (2004, in preparation) in M101. 
Given the higher spatial resolution, the HST photometry samples
bound stellar clusters, while our GALEX census samples the star
forming complexes  on spatial scales of $\lesssim$ 160~pc. 
For M51 we do not perform a similar analysis given the smaller number
of compact sources detected, and distortion of the images
which limits the accuracy of compact sources photometry.  

In summary,  the GALEX images provided a complete census of young,
 compact star-forming complexes. 
The UV bands are extremely sensitive to young stellar populations. 
E.g., the [intrinsic] \fuvmag-$r$ ($\approx$ \fuvmag-V) 
colors of clusters (SSP)  with
ages  of 1, 10, 100 Myrs are -1.78, -0.30, 1.11 (AB mag), while the
$ u - g$ ($ g - r$) [approximately  
B-V (U-B)]  colors are  -0.49, 0.06, -0.03  (-0.39, 0.01, 0.69). 
 Thus by including GALEX measurements we gain  a finer resolution
for parameters of  young associations and complexes, hence a clearer, unbiased 
census of  recent star formation.  Figure \ref{fage} 
indicates that SF in M101 occured 
over the last 10$^9$ yrs,
both in stellar clusters (Chandar et al. 2004) and in larger
complexes (GALEX results, this paper). 
This analysis will be extended to other resolved galaxies in the
GALEX NGS, to  compare  their cluster systems and 
star-formation histories.

\section{The properties of the stellar populations}
\label{sprofiles}

\begin{figure}
\epsscale{1.0}
\plotone{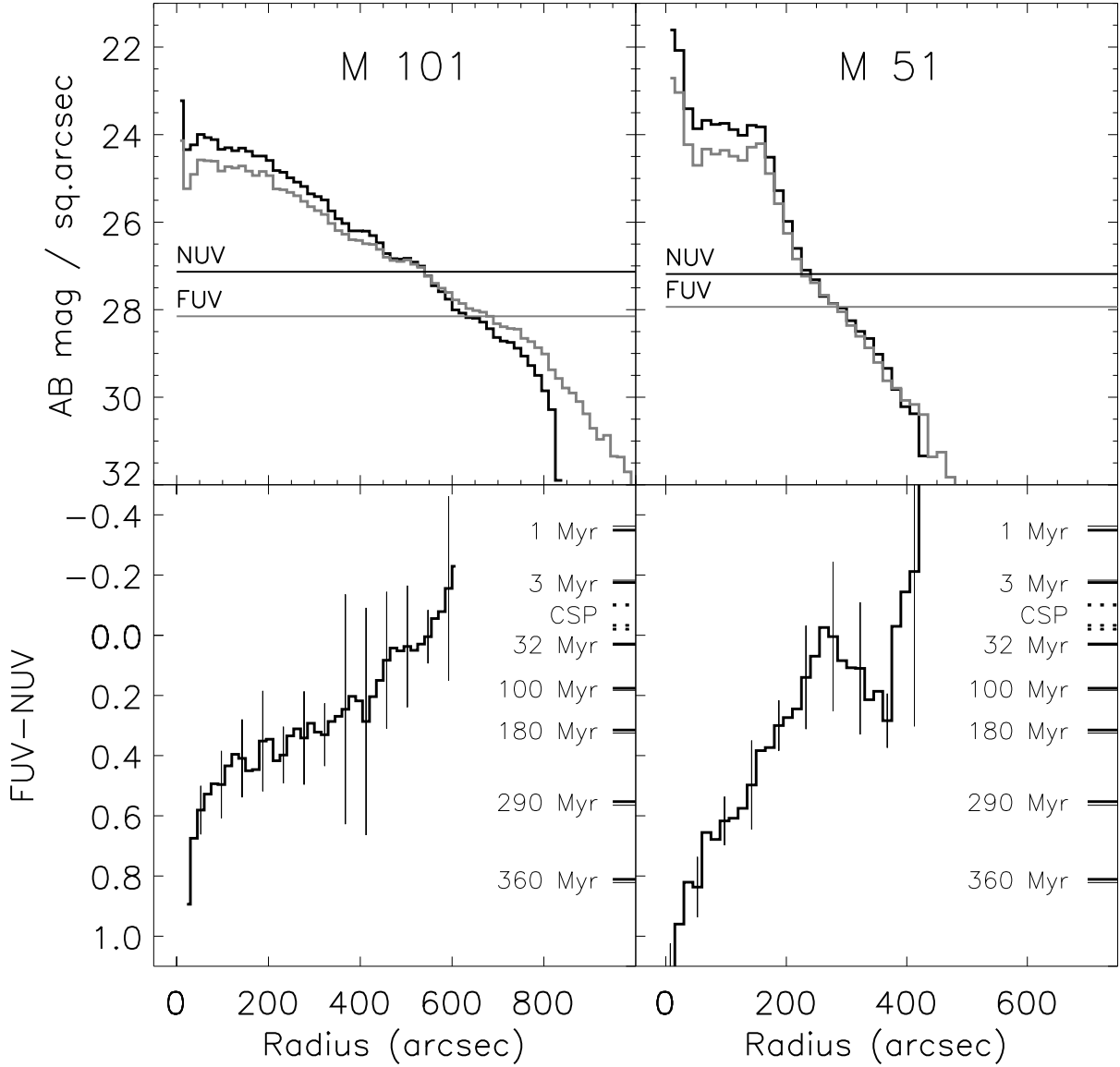} 
\vskip -3.0cm
\caption{Radial profiles of \fuvmag and \nuvmag ~ median surface brightness, 
and FUV-NUV color. 
Errors ($\pm$ 1 $\sigma$) are shown for representative points of the FUV-NUV color.
For the brightness profiles,  errors range between 0.075 to 0.3 mag
in the range 24-29 mag. 
The horizontal lines are the estimated sky surface brightness in each band. 
On the  \fuvmag-\nuvmag panels, we show (line segments)
theoretical \fuvmag-\nuvmag  values for 
SSP models of ages 1, 3, 32, 100, 180, 290, 360 Myr, 
and CSP models of ages 0.1, 1, 10 Gyr (dashed segments). 
\label{fradprof} }
\end{figure}

To explore the average properties of the stellar population as
a function of galactocentric distance, we computed 
radial profiles of the \fuvmag and \nuvmag median surface brightness
for each galaxy  within concentric
elliptical annuli, oriented in accordance with the galaxy inclination
and position angle (M101: i=18$^o$, PA=39$^o$, 
Bosma et al. 1981; M51: i=20$^o$, PA=170$^o$, Rots et al. 1990).
 Within each annulus we measured the median surface brightness (thus removing 
the discrete peaked sources, and the foreground stars)
and subtracted an average sky background,  determined
in a wide elliptical annulus exterior to the galaxy.  
In Figure \ref{fradprof} we  plot the
surface brightness as a function of radius in the two GALEX 
bands, as well as the \fuvmag-\nuvmag color. 
Values of this color for SSP and CSP models are indicated for a range of ages.
 As shown previously,  this color is essentially
reddening-free, thus gives a direct indication of the age.  
Observed colors are redder than CSP models for a wide range of ages,
across most of the galaxy, excluding this scenario.   SSP 
models represent the extreme opposite of the CSP scenario, and SSP
ages have only indicative value, since we are comparing colors of
azimuthally averaged populations, 
and single bursts of star formation (clusters, associations) would likely
 be localized. 
The SSP ages indicate a  more recent star-formation activity 
towards outer regions (spanning two dex in age) for M101, and much less
conspicuous fluctuations in M51, while the bulge has an  older (average) age.
The comparison of Figure \ref{fradprof} with the FUV 
radial profile  from  UIT data 
 (Kuchinski et al. 2000) shows the advantage of the GALEX measurements.
In addition to providing the \fuvmag-\nuvmag color,
for both M101 and M51 the profiles in Figure \ref{fradprof} extend to larger
galactocentric radii, sampling outer disk areas, about 3
magnitudes fainter than the UIT data (note again that we use 
AB magnitudes, which differ from the UIT magnitude system
- for \fuvmag - by $\approx$ 2.8  mag).

The radial profiles of the \fuvmag-\nuvmag color 
in the Local Group galaxies M~33 and M~31, and in M83, 
(Thilker et al, 2004a,b this issue) show the same trend of the 
\fuvmag-\nuvmag color becoming bluer outwards,
 the magnitude of the color gradient 
differing  from galaxy to galaxy. An extensive comparison among a
representative sample of nearby galaxies will follow in a forthcoming paper.

\begin{acknowledgments}
Acknowledgement: 

GALEX (Galaxy Evolution Explorer) is a NASA Small Explorer, launched in April 2003.
We gratefully acknowledge NASA's support for construction, operation,
and science analysis of the GALEX mission,
developed in cooperation with the Centre National d'Etudes Spatiales
of France and the Korean Ministry of 
Science and Technology. 
 We are very grateful to R. Chandar for helpful 
discussions, and to the referee for many useful suggestions.  
\end{acknowledgments}

\newpage

\newpage
\renewcommand{\arraystretch}{.6}
{\footnotesize
\begin{deluxetable}{lccclccc}
\tablewidth{0pt}
\tablecaption{The sample of sources in the different bands {\label{tlimits}} }
\tablehead{
\multicolumn{1}{l}{Band}    & \multicolumn{1}{c}{Sources}    & \multicolumn{1}{c}{Mag.lim.} &   
\multicolumn{1}{c}{Sources}   & \multicolumn{1}{c}{Mag.lim.}    & 
\multicolumn{1}{c}{Sources}   & \multicolumn{1}{c}{Mag.lim.}    \\
\multicolumn{1}{l}{ }            &    \multicolumn{2}{c}{(error$<$0.2 mag)} 
                     & \multicolumn{2}{c}{(error$<$0.1 mag)}  & 
   \multicolumn{2}{c}{(error$<$0.05 mag)}       
}
\startdata
\multicolumn{5}{l}{\it M~101 \rm } \\
\fuvmag &  1034 &   22.1 &     383  &   20.8  &  82    &  19.1 \\
\nuvmag  &  2179 &   22.8 &     898 &   21.7  &  249   &  20.4\\
u     &  1351 &   22.4 &     878 &   21.7  &  389   &  20.6\\
g     &  1315 &   23.0 &     939 &   22.3  &  539   &  21.4\\
r     &  1233 &   22.6 &     851 &   22.0  &  467   &  21.2\\
i     &  1066 &   22.4 &     715  &   21.6  &  419   &  20.4\\
z     &  796  &   21.1 &     468  &   20.4  &  248   &  19.1\\
\multicolumn{5}{l}{\it M~51\rm } \\
\fuvmag              & 200  &    22.5  &      78  &      21.1  &  25 &      19.6\\
\nuvmag           & 622  &    22.9  &      257 &      21.9  &  83 &     20.7\\
u                 & 268  &    ---  &      140 &     21.8   &  54 &      20.7\\
g                 & 253  &    ---  &      153 &     22.5   &  76 &     21.6\\
r                 & 228  &    ---  &      131 &     21.9   &  71 &      21.0\\
i                 & 205  &    ---  &      115 &     21.7   &  65 &      20.6\\
z                 & 170  &    ---  &      85  &     20.2   &  42 &      19.3\\
\enddata
\end{deluxetable}
}
\normalsize


\end{document}